\documentclass{bsl}

\usepackage{amsmath,amsthm,amssymb}

\usepackage{stmaryrd}
\usepackage{color,soul,coloring}

\usepackage[pdftex]{graphicx} 
\usepackage{latexsym,url}

\theoremstyle{plain}

\theoremstyle{definition}

\theoremstyle{remark}

\begin{document}
\AuthorTitle{Christoph Benzm\"uller and David Fuenmayor}{Computer-supported Analysis of Positive Properties,
  Ultrafilters and Modal Collapse in Variants of G\"odel's Ontological Argument}
\begin{abstract}
  Three variants of Kurt G\"odel's ontological argument, proposed by
  Dana Scott, C. Anthony Anderson and Melvin Fitting, are encoded and
  rigorously assessed on the computer.  
  In contrast to Scott's version of G\"odel's argument the two variants
  contributed by Anderson and Fitting avoid modal collapse. Although
  they appear quite different on a cursory reading they are in fact
  closely related. This has been revealed in the computer-supported
  formal analysis presented in this article. Key
  to our formal analysis is the utilization of suitably
  adapted notions of (modal) ultrafilters, and a careful distinction
  between extensions and intensions of positive properties.
\end{abstract}

\section{Introduction}
The premises of the variant of the modal ontological
argument \cite{GoedelNotes} which was found in Kurt G\"odel's ``Nachlass'' 
are inconsistent; this holds already in base modal logic \textbf{K}
\cite{C55,C40}. The premises of Scott's \cite{ScottNotes} variant of G\"odel's work, in
contrast, are  consistent \cite{C40,C55}, but they
imply the modal collapse, $\varphi \rightarrow \Box \varphi$, which has
by many philosophers been considered an undesirable side effect;
cf.~Sobel \cite{sobel2004logic} and the references
therein.\footnote{The modal collapse was already noted by Sobel
  \cite{Sobel,sobel2004logic}. One might conclude from it, that the
premises of G\"odel's argument imply that everything is determined, or
alternatively, that there is no free will. Sre\'{c}ko Kova$\check{\text{c}}$ \cite{Kovacs2012}
  argues that modal collapse was eventually intended by G\"odel.}

In this article we formally encode and analyze, starting with Scott's
variant, two prominent further emendations of G\"odel's work both of which
successfully avoid modal collapse. These two variants have been
contributed by
C.~Anthony Anderson \cite{Anderson90,AndersonGettings} and Melvin
Fitting \cite{fitting02:_types_tableaus_god}, and on a cursory reading
they appear quite different.  Our formal analysis, however, shows that
from a certain mathematical perspective they are in fact closely
related.

Two notions are particularly important in our analysis. From set
theory, resp.~topology, we borrow and suitably adapt, for use in our modal
logic context, the notion of ultrafilter and apply it in two
different versions to the set of positive properties. From the
philosophy of language we adopt the distinction between intensions and
extensions of (positive) properties. Such a distinction has been
suggested already by Fitting in his book \emph{``Types, Tableaus
  and G\"odel's God''} \cite{fitting02:_types_tableaus_god}, which we
take as a starting point in our formalization work.

Utilizing these notions, and extending Fitting's analysis, the
modifications as introduced by Anderson and Fitting to G\"odel's concept
of positive properties are formally studied and compared. Our
computer-supported analysis, which is carried out in the proof assistant
system Isabelle/HOL~\cite{Isabelle2}, is technically enabled by the
universal logical reasoning approach \cite{J41}, which exploits
shallow semantical embeddings (SSEs) of various logics of
interest---such as intensional higher-order modal logics (IHOML) in
the present article---in Church's simple type theory~\cite{J43}, aka. classical
higher-order logic (HOL). This approach enables the reuse of existing,
interactive and automated, theorem proving technology for HOL to
mechanize also non-classical higher-order reasoning.

Some of the findings reported in this article have, at an abstract
level, already been summarized in the literature before \cite{J47,C74,C65}, but they
have not been published in full detail yet (for example, the notions of ``modal'' ultrafilters,
as employed in our analysis, have not been made precise in these
papers). This is the contribution of this article.

 In fact, we present and explain in detail the SSE of
intensional higher-order modal logic (IHOML) in HOL (\S\ref{IHOML}),
the encoding of different types of modal filters and modal ultrafilters in
HOL (\S\ref{Ultrafilters}), and finally the encoding and analysis of the three mentioned
variants of G\"odel's ontological argument in HOL utilizing the SSE
approach (\S\ref{Scott}, \S\ref{Anderson} and \S\ref{Fitting}). We
start out (\S\ref{SSE}) by pointing to related prior work and by outlining the
SSE approach.

\section{Prior Work and the SSE Approach} \label{SSE} 
The key ideas of the shallow semantical embedding
(SSE) approach, as relevant for the remainder of this article,
are briefly outlined.
This section is intended to make the article sufficiently
self-contained and to give references to related prior work.  The
presentation in this section is taken and adapted from a recently
published related article~\cite[\S 1.1]{J47}; readers already familiar with the SSE
approach may simply skip it, and those who need further details may 
consult further related documents~\cite{J23,J41}.

Earlier papers, cf.~\cite{J41} and the references therein, focused on
the development of SSEs. These papers show that the standard
translation from propositional modal logic to first-order logic can be
concisely modeled (i.e., embedded) within higher-order theorem
provers, so that the modal operator $\Box$, for example, can be
explicitly defined by the $\lambda$-term
$\lambda \varphi. \lambda w. \forall v. (R w v \rightarrow \varphi
v)$,
where $R$ denotes the accessibility relation associated with $\Box$.
Then one can construct first-order formulas involving $\Box\varphi$
and use them to represent and proof theorems. Thus, in an SSE, the
target logic is internally represented using higher-order constructs
in a proof assistant system such as Isabelle/HOL.  The first author, in
collaboration with Paulson \cite{J23}, developed an SSE that captures
quantified extensions of modal logic (and other non-classical
logics). For example, if $\forall x. \phi x$ is shorthand in higher-order logic (HOL) for
$\Pi (\lambda x. \phi x)$, then $\Box \forall x Px$ would be
represented as $\Box \Pi' (\lambda x. \lambda w. P x w)$, where $\Pi'$
stands for the $\lambda$-term
$\lambda \Phi . \lambda w . \Pi(\lambda x . \Phi x w)$, and the $\Box$
gets resolved as described above.

To see how
  these expressions can be resolved to produce the right
  representation, consider the following series of reductions:\\
  \hspace*{.2in}\begin{tabular}{lll} $\Box\forall x P x$ & $\equiv$ &
    $\Box \Pi' (\lambda x. \lambda w. P x w)$\\ & $\equiv$ &
    $\Box ((\lambda \Phi . \lambda w . \Pi(\lambda x . \Phi x w))
    (\lambda x. \lambda w. P x w))$\\
    & $\equiv$ &
    $\Box (\lambda w . \Pi(\lambda x . (\lambda x. \lambda w. P x w) x
    w))$\\
    & $\equiv$ & $\Box (\lambda w . \Pi(\lambda x . P x w))$\\ &
    $\equiv$ &
    $(\lambda \varphi. \lambda w. \forall v. (R w v \rightarrow
    \varphi v)) (\lambda w . \Pi(\lambda x . P x w))$\\
    & $\equiv$ &
    $(\lambda \varphi. \lambda w. \Pi (\lambda v . R w v \rightarrow
    \varphi v)) (\lambda w . \Pi(\lambda x . P x w))$\\
    & $\equiv$ &
    $(\lambda w. \Pi (\lambda v . R w v \rightarrow (\lambda w
    . \Pi(\lambda x . P x w)) v)) $\\
    & $\equiv$ &
    $(\lambda w. \Pi (\lambda v . R w v \rightarrow \Pi(\lambda x . P
    x v)) ) $\\
    & $\equiv$ &
    $(\lambda w. \forall v . R w v \rightarrow \forall x . P x v) $\\
    & $\equiv$ & $(\lambda w. \forall v x . R w v \rightarrow P x v) $
  \end{tabular}\\
Thus, we end up with a representation of  $\Box\forall x P x$ in
HOL.  Of course, types are assigned to each term of the HOL language. 
More precisely, in the SSE presented in 
Fig.~\ref{fig:IHOML}, we will assign individual terms (such as variable $x$
above) the type \texttt{e}, and terms denoting worlds (such as
variable $w$ above) the type \texttt{i}. From such base
choices, all other types in the above presentation can be inferred.
While types have been
omitted above, they will often be given in the remainder of this article.

The SSE technique provided a fruitful starting point for a natural
encoding of G\"odel's ontological argument in second-order modal
logics \textbf{S5} and \textbf{KB}~\cite{C40}.  Initial studies
investigated G\"odel's and Scott's variants of the argument within the
higher-order automated theorem prover (henceforth ATP)
LEO-II~\cite{J30}.  Subsequent work deepened these assessment
studies~\cite{C55,C60}. Instead of using LEO-II, these studies
utilized the higher-order proof assistant Isabelle/HOL, which is
interactive and which also supports strong proof automation. Some
experiments were also conducted with the proof assistant Coq
\cite{C44}. Further work (see the references in~\cite{J47,J41})
contributed a range of similar studies on variants of the modal
ontological argument that have been proposed by
Anderson~\cite{Anderson90}, Anderson and
Gettings~\cite{AndersonGettings}, H\'ajek~\cite{Hajek1,Hajek2,Hajek3},
Fitting~\cite{fitting02:_types_tableaus_god}, and
Lowe~\cite{Lowe2013}.  Particularly relevant for this article is some
prior formalization work by the authors that has been presented
in~\cite{J35,C65}.  The use of ultrafilters to study the distinction
between extensional and intensional positive properties in the
variants of Scott, Anderson and Fitting has first been mentioned in
invited keynotes presented at the AISSQ-2018~\cite{C74} and
the FMSPh-2019~\cite{R74} conferences.

\section{Further Preliminaries} 
The formal analysis in this article takes Fitting's book
\cite{fitting02:_types_tableaus_god} as a starting point; see
also~\cite{J35,C65}. Fitting suggests to carefully distinguish between
intensions and extensions of positive properties in the context of
G\"odel's argument, and, in order to do so within a single framework,
he introduces a sufficiently expressive higher-order modal logic
enhanced with means for the explicit representation of intensional
terms and their extensions, which we have termed intensional
higher-order modal logic (IHOML) in previous work \cite{C65}. The SSE
of IHOML in HOL, that we utilize in the remainder of this article, is
presented in~\S\ref{IHOML}. Notions of ultrafilters on sets of
intensions, resp.~extensions, of (positive) properties are then
introduced in \S\ref{Ultrafilters}. Since we develop, explain and
discuss our formal encodings directly in Isabelle/HOL
\cite{Isabelle2}, some familiarity with this proof assistant and its
background logic HOL \cite{J43} is assumed.

\subsection{Intensional Higher-Order Modal Logic in HOL} \label{IHOML}
\begin{figure}[tp] 
\includegraphics[width=1\textwidth]{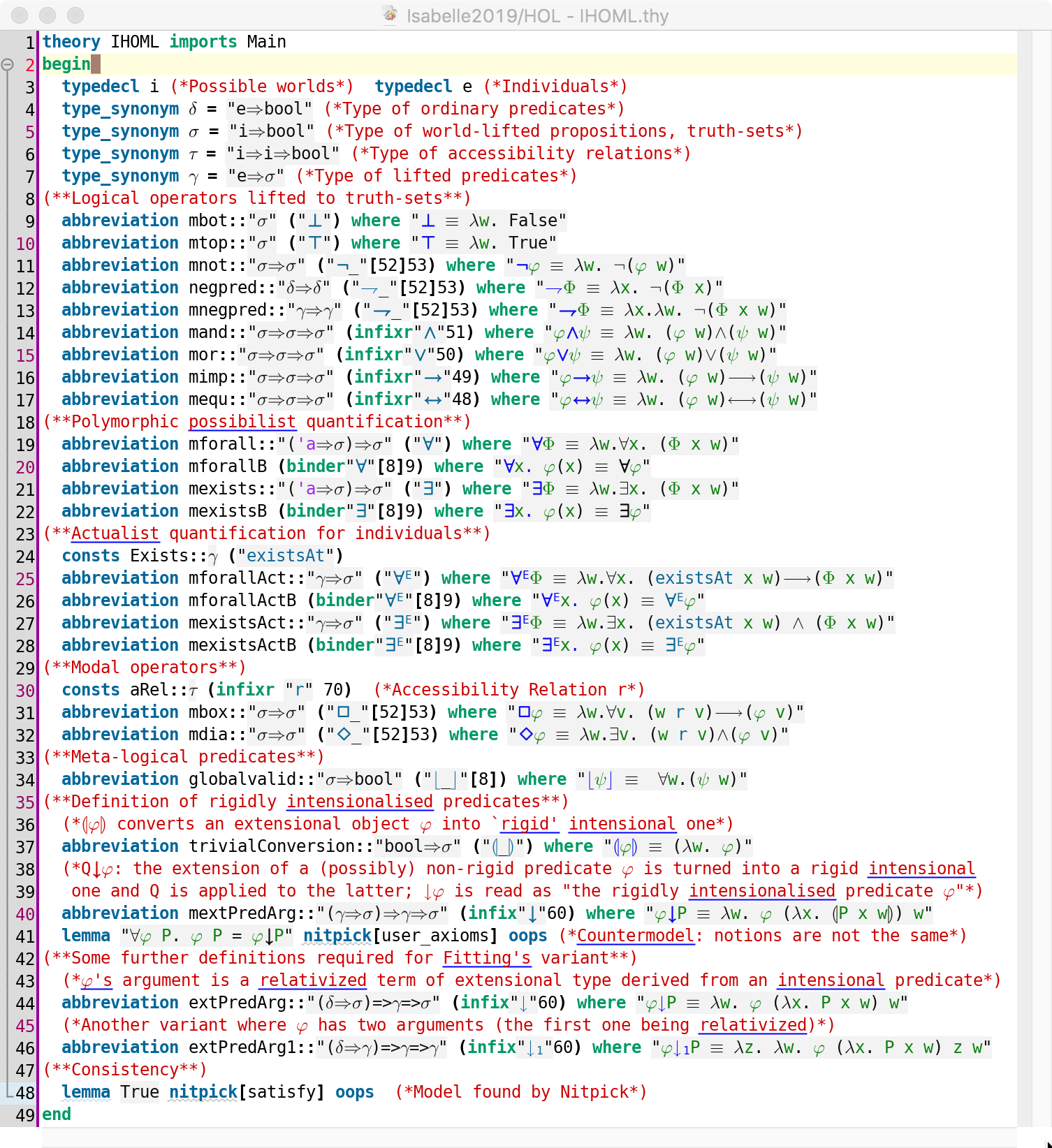}
\caption{Shallow semantical embedding of IHOML in HOL. \label{fig:IHOML}}
\end{figure}

An encoding of IHOML in Isabelle/HOL utilizing the SSE approach, is
presented in Fig.~\ref{fig:IHOML}. It starts in line 3 with the
declaration of two base types in HOL as mentioned before: type
\texttt{i} stands for possible worlds and type \texttt{e} for
entities/individuals. To keep the encoding concise some type synonyms
are introduced in lines 4--7, which we explain next.

$\delta$ and $\sigma$ abbreviate the types of predicates
$\texttt{e}{\Rightarrow}\texttt{bool}$ and
$\texttt{i}{\Rightarrow}\texttt{bool}$, respectively. Terms of 
type $\delta$ represent (extensional) properties of individuals.
Terms of type $\sigma$  can be seen to represent world-lifted propositions,
i.e., truth-sets in Kripke's modal relational
semantics~\cite{sep-logic-modal}. Note that the explicit transition
from modal propositions to terms (truth-sets) of type
$\sigma$ is a key aspect in SSE approach; see the literature
\cite{J41} for further details. In the remainder of this article we
make use of phrases such as ``world-lifted'' or ``$\sigma$-type'' terms
to emphasize this conversion in the SSE approach.

$\tau$, which abbreviates the type
$\texttt{i}{\Rightarrow} \texttt{i}{\Rightarrow}\texttt{bool}$, stands for the type of accessibility relations in modal relational
semantics, and $\gamma$, which stands for $\texttt{e}{\Rightarrow}\sigma$,
is the type of world-lifted, intensional properties.

In lines 8--32 in Fig.~\ref{fig:IHOML} the modal logic connectives are introduced. For example,
in line 15 we find the definition of the world-lifted
$\boldsymbol\vee$-connective (which is of type $\sigma{\Rightarrow}
\sigma{\Rightarrow} \sigma$; type information is given here explicitly after the
\texttt{::}-token for `\texttt{mor}', which is the
ASCII-denominator for the infix-operator $\boldsymbol\vee$ as introduced in
parenthesis shortly after). $\varphi_\sigma \boldsymbol\vee \psi_\sigma$ is defined
as abbreviation for the truth-set $\lambda w_i.  \varphi_\sigma
w_i \vee \psi_\sigma w_i$ (i.e., $\boldsymbol\vee$ is associated with the
lambda-term $\lambda \varphi_\sigma. \lambda \psi_\sigma. \lambda w_i.  \varphi_\sigma
w_i \vee \psi_\sigma w_i$). In the remainder we generally use bold-face
symbols for world-lifted connectives (such as $\boldsymbol\vee$) in
order to rigorously distinguish them from their ordinary
counterparts (such as $\vee$) in the meta-logic HOL.

The world-lifted $\boldsymbol\neg$-connective is introduced in line
11, $\boldsymbol\bot$ and $\boldsymbol\top$ in lines 9--10, and
respective further abbreviations for conjunction, implication and
equivalence are given in lines 14, 16 and 17, respectively.  The
operators $\boldsymbol\rightharpoondown$ and $\rightharpoondown$,
introduced in lines 12 and 13, negate properties of types
$\delta$ and $\gamma$, respectively; these operations occur
in the premises in the works of Scott, Anderson and Fitting which govern the definition
of positive properties.

As we see in Fig.~\ref{fig:IHOML}, types can often be omitted in Isabelle/HOL due to
the system's internal type inference mechanism. This feature is
exploited in our formalization to some extend to improve
readability. However, for all \emph{new} abbreviations and
definitions, we always explicitly declare the types of the freshly
introduced symbols; this not only supports a better intuitive
understanding of these notions but also reduces the number of
polymorphic terms in the formalization (since polymorphism may generally
cause decreased proof automation performance).

The world-lifted modal $\boldsymbol\Box$-operator and the
polymorphic, world-lifted universal quantifier $\boldsymbol\forall$, as already
discussed in \S\ref{SSE}, are introduced in lines 31 and 19,
respectively (the \texttt{'a} in the type declaration for $\boldsymbol\forall$ represents a type
  variable). In line 20, user-friendly binder-notation for $\boldsymbol\forall$
is additionally defined. In addition to the (polymorphic) possibilist
quantifiers, $\boldsymbol\forall$ and $\boldsymbol\exists$, defined this way in lines 19--22, further
actualist quantifiers, $\boldsymbol\forall^E$ and
$\boldsymbol\exists^E$, are introduced in
lines 24--28; their definition is guarded by an explicit, possibly empty,
\texttt{existsAt} predicate, which encodes whether an individual
object actually ``exists''
at a particular given world, or not. These additional actualist quantifiers
are declared 
non-polymorphic, so that they support quantification over
individuals only. In the subsequent analysis of the variants of
G\"odel's argument, as contributed by Scott,
Anderson and Fitting, we will indeed use $\boldsymbol\forall$ and $\boldsymbol\exists$ for
different types in the type hierarchy of HOL, while keeping $\boldsymbol\forall^E$
and $\boldsymbol\exists^E$ for quantification over individuals only.
 
The notion of global validity of a world-lifted formula $\psi_\sigma$,
denoted as $\lfloor\psi\rfloor$, is introduced in line 34 as an
abbreviation for $\forall w_\texttt{i}.  \psi w$.

Note that an (intensional) base modal logic \textbf{K} is introduced in the theory
\texttt{IHOML} (Fig.~\ref{fig:IHOML}). In later sections we will switch to logics \textbf{KB} and \textbf{S5} by
postulating respective conditions (symmetry, and additionally
reflexivity and transitivity) on the accessibility relation  \texttt{r}.

In lines 35--46 some further abbreviations are declared, which address
the mediation between intensions and extensions of
properties. World-lifted propositions and intensional properties are
modeled as terms of types $\sigma$ and $\gamma$
respectively, i.e., they are technically handled in HOL as functions
over worlds whose extensions are obtained by applying them to a given
world $w$ in context. The operation
$\boldsymbol\llparenthesis \varphi \boldsymbol\rrparenthesis$ in line
37 is trivially converting a world-independent proposition of Boolean
type into a \emph{rigid} world-lifted proposition of type $\sigma$; the
rigid world-lifted propositions obtained from this trivial conversion
have identical evaluations in all worlds.

The ${\boldsymbol\downarrow}$-operator in line 40, which is of type
${(\gamma{\Rightarrow}\sigma){\Rightarrow}\gamma{\Rightarrow}\sigma}$,
is slightly more involved. It evaluates its second argument, which is
a property $P$ of type $\gamma$, for a given world $w$, and it then \emph{rigidly
  intensionalizes} the obtained extension of $P$ in $w$. For
technical reasons, however, $\boldsymbol\downarrow$ is introduced as a
binary operator, with its first argument being a world-lifted
predicate $\varphi_{\gamma{\Rightarrow}\sigma}$ that is being applied
to the rigidly intensionalized ${\boldsymbol\downarrow}P_\gamma$; in
fact, all occurrences of the $\boldsymbol\downarrow$-operator in our
subsequent sections will have this binary pattern.

The lemma statement in line 41 confirms that intensional properties
$P_\gamma$ are generally different from their rigidly intensionalized
counterparts ${\boldsymbol\downarrow}P_\gamma$: Isabelle/HOL's model
finder Nitpick \cite{Nitpick} generates a countermodel to the claim
that they are (Leibniz-)equal.

A related (non-bold) binary operator $\downarrow$, of type
${(\delta{\Rightarrow}\sigma){\Rightarrow}\gamma{\Rightarrow}\sigma}$,
is introduced in line 44. Its first argument is a predicate
$\varphi_{\delta{\Rightarrow}\sigma}$ applicable to extensions of
properties, and its second argument is an intensional property. The
$\downarrow$-operator evaluates its second argument $P_\gamma$ in a
given world $w$, thereby obtaining an extension ${\downarrow}P_\gamma$
of type $\delta$, and then it applies its first argument
$\varphi_{\delta{\Rightarrow}\sigma}$ to this extension.  The
$\downarrow_1$-operator is analogous, but its first argument
$\varphi$ is now of type ${\delta{\Rightarrow}\gamma}$, which can be
understood as world-lifted binary predicate whose first argument is of
type $\delta$ and its second argument of type $\texttt{e}$. The
$\downarrow_1$-operator evaluates the intensional argument
$P_\gamma$, given to it in second position, in a given world $w$, and
it then applies $\varphi_{\delta{\Rightarrow}(e{\Rightarrow}\sigma)}$
to the result of this operation and subsequently to its (unmodified)
second argument $z_e$.

In line 48, consistency of the introduced concepts is confirmed by the
model finder Nitpick \cite{Nitpick}. Since only
abbreviations and no axioms have been introduced so far, the consistency of the Isabelle/HOL
theory \texttt{IHOML} in Fig.~\ref{IHOML} is actually evident.

\subsection{Filters and Ultrafilters} \label{Ultrafilters} 
\begin{figure}[tp] \centering
\includegraphics[width=1\textwidth]{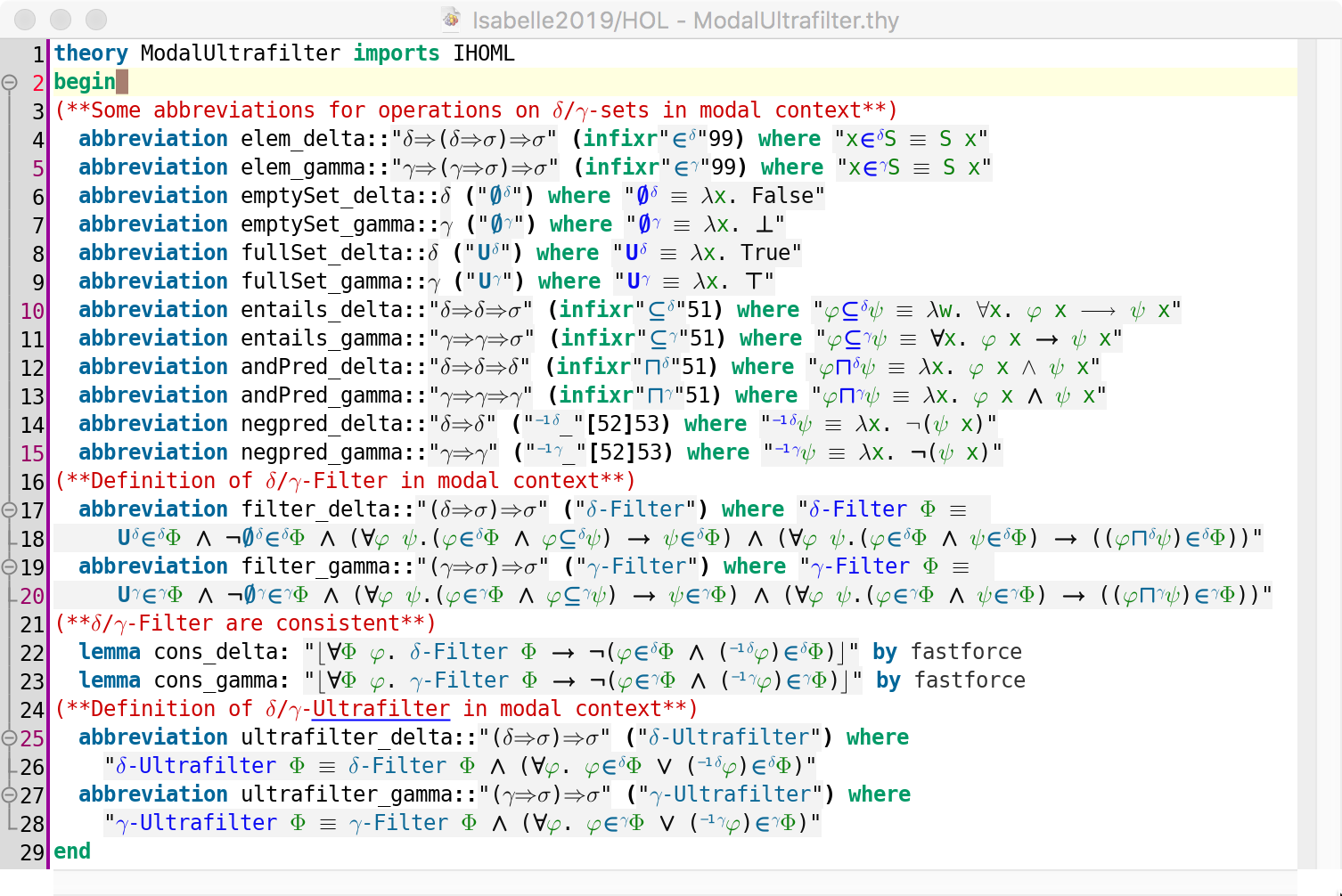}
\caption{Definition of $\delta/\gamma$-Filters and $\delta/\gamma$-Ultrafilters. \label{fig:Ultrafilters}}
\end{figure}
Two related world-lifted notions of modal filters and modal
ultrafilters are defined in
Fig.~\ref{fig:Ultrafilters}; for a general introduction to filters
and ultrafilters we refer to the corresponding mathematical literature (e.g.~\cite{UNIALGEBRA}).

$\delta$-Ultrafilters are introduced in line 26 as world-lifted
characteristic functions of type
$(\delta{\Rightarrow}\sigma){\Rightarrow}\sigma$. They thus denote
$\sigma$-sets of $\sigma$-sets of objects of type $\delta$.  In other
words, a $\delta$-Ultrafilter is a $\sigma$-subset of the
$\sigma$-powerset of $\delta$-type property extensions.

A $\delta$-Ultrafilter $\phi$ is defined as a $\delta$-Filter
satisfying an additional maximality condition:
$\forall \varphi.  \varphi\boldsymbol\in^\delta\phi \boldsymbol\vee
(^{-1\delta}\varphi)\boldsymbol\in^\delta\phi$,
where $\boldsymbol\in^\delta$ is elementhood of $\delta$-type objects
in $\sigma$-sets of $\delta$-type objects (see line 4), and where
$^{-1\delta}$ is the relative set complement operation on sets of
entities (see line 14).

The notion of $\delta$-Filter is introduced in lines 17
and 18.  A $\delta$-Filter $\phi$ is required to
\begin{itemize} 
\item be large: $\textbf{U}^\delta\boldsymbol\in^\delta\phi$, where $\textbf{U}^\delta$ denotes
the full set of $\delta$-type objects we start with (see line 8), 
\item exclude the empty set:
  $\boldsymbol\emptyset^\delta\boldsymbol\not\boldsymbol\in^\delta\phi$,
  where $\boldsymbol\emptyset^\delta$ is the world-lifted empty set of
  $\delta$-type objects (see line 6),
\item be closed under supersets:
  $\boldsymbol\forall \varphi\,
  \psi. (\varphi\boldsymbol\in^\delta\phi \boldsymbol\wedge
  \varphi\boldsymbol\subseteq^\delta\psi ) \boldsymbol\rightarrow
  \psi\boldsymbol\in^\delta\phi$
  (the world-lifted subset relation $\boldsymbol\subseteq^\delta$ is
  defined in line 10), and
\item be closed under intersections:
  $\boldsymbol\forall \varphi\,
  \psi. (\varphi\boldsymbol\in^\delta\phi \boldsymbol\wedge
  \psi\boldsymbol\in^\delta\phi) \boldsymbol\rightarrow
  (\varphi\boldsymbol\sqcap^\delta\psi)\boldsymbol\in^\delta\phi$
  (the intersection operation $\boldsymbol\sqcap^\delta$
  is defined in line 12).
\end{itemize}

$\gamma$-Ultrafilters, which are of type
$(\gamma{\Rightarrow}\sigma){\Rightarrow}\sigma$, are analogously
defined as a $\sigma$-subset of the $\sigma$-powerset
of $\gamma$-type property extensions.

The distinction of both notions of ultrafilters is needed in our subsequent
investigation. This is because we will rigorously distinguish between 
positive property intensions (as used by Scott and Anderson) and
positive property extensions (as utilized by Fitting). 

By using polymorphic definitions, several ``duplications'' of abbreviations in
the theory \texttt{ModalUltrafilter} (Fig.~\ref{fig:Ultrafilters}) could be avoided. To support a
more precise understanding of $\delta$- and $\gamma$-Ultrafilters, and
their differences, however, we have decided to be very transparent and
explicit regarding type information in the provided definitions.

\section{Scott's Variant of G\"odel's Argument} \label{Scott} 
\begin{figure}[tp] \centering \footnotesize
\fcolorbox{white}{gray!10}{
\begin{tabular}{ll} 
\multicolumn{2}{c}{\textbf{Scott's Axioms and Definitions}} \\ \hline  \\[-.5em]
(df.$G$) & $\texttt{G}\, x \equiv \forall Y_\gamma. \mathcal{P}\, Y
  \rightarrow Y x$ \\ 
(A1a) & $\forall X. \mathcal{P}({\rightharpoondown} X)  \rightarrow \neg (
  \mathcal{P}\, X)$     \qquad {where $\rightharpoondown$ is
    set/predicate negation} \\
(A1b) & $\forall X. \neg ( \mathcal{P}\, X)\rightarrow
        \mathcal{P}({\rightharpoondown} X)$ \\
(A2) & $\forall X Y. (\mathcal{P}\, X \wedge \Box  (\forall^E
  z. X z \rightarrow Y z)) 
  \rightarrow  \mathcal{P}\, Y$ \\
(A3)  & $\forall Z X. ((\forall Y. Z\,Y \rightarrow \mathcal{P}\, Y)\wedge \Box (\forall x. X
  x \leftrightarrow (\forall Y. Z\,Y \rightarrow Y x))) 
  \rightarrow  \mathcal{P} X$ \\
(A4) & $\forall X. \mathcal{P}\, X \rightarrow  \Box (\mathcal{P}\,
  X)$ \\
(df.$\mathcal{E}$) & $\mathcal{E}\, Y x \equiv Y x \wedge (\forall
  Z. Z x \rightarrow \Box  (\forall^E
  z. Y z \rightarrow Z z))$ \\
(df.$\text{NE}$) & $\text{NE } x \equiv \forall Y.  \mathcal{E}\, Y
  x \rightarrow \Box \exists^E\, Y$ \\
(A5) &  $\mathcal{P}\, \text{NE}$ \\ 
\end{tabular}
}
\vskip1em
\includegraphics[width=1\textwidth]{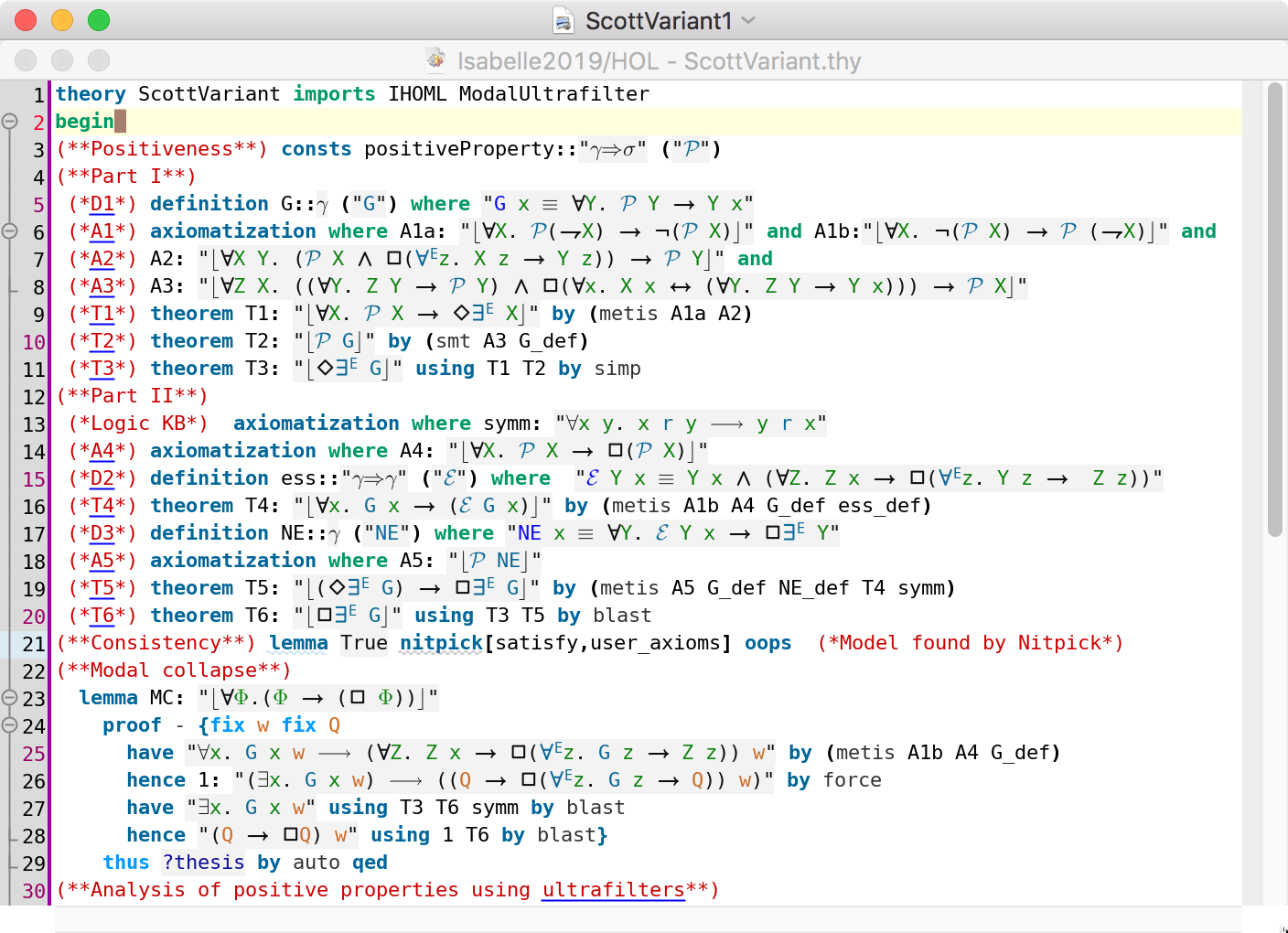}
\caption{Scott's variant of G\"odel's argument, following Fitting \cite{fitting02:_types_tableaus_god}. \label{fig:Scott1}}
\end{figure}

\begin{figure}[tp] \centering
\includegraphics[width=1\textwidth]{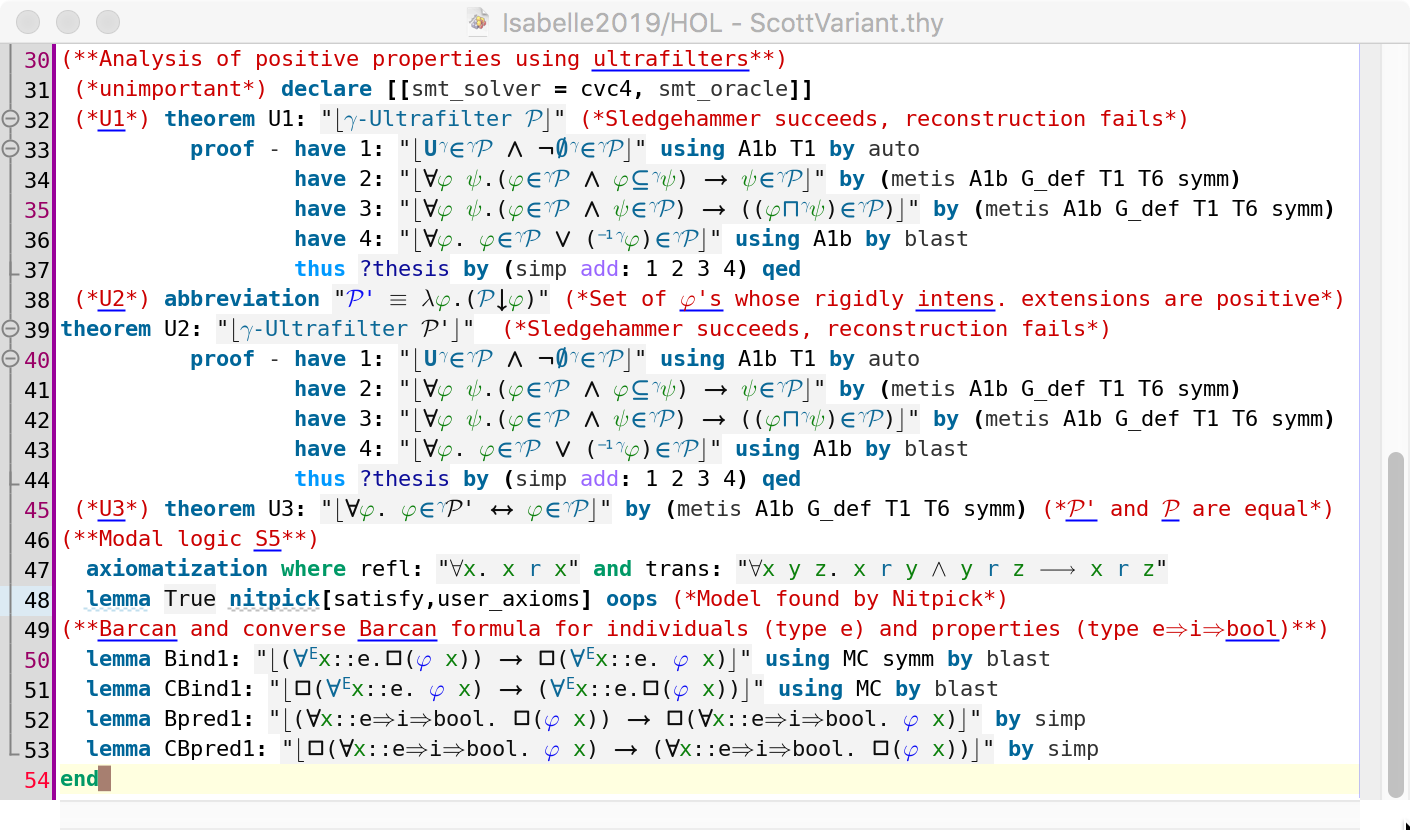}
\caption{Ultrafilter-analysis of Scott's variant (continued from~Fig.~\ref{fig:Scott1}). \label{fig:Scott2}}
\end{figure}

Scott's
variant of G\"odel's argument has been reproduced by Fitting in his
book \cite{fitting02:_types_tableaus_god}. It is Fitting's
formalization of Scott's variant that we have encoded and verified
first in our computer-supported analysis of positive properties,
ultrafilters and modal collapse.  This encoding of
Scott's variant is presented in Fig.~\ref{fig:Scott1} and its
presentation is continued in
Fig.~\ref{fig:Scott2}. 

Part I of the argument is reconstructed in lines 4-11 of
Fig.~\ref{fig:Scott1} and verified with automated reasoning
tools.\footnote{The automated reasoning tools that are integrated with
  Isabelle/HOL, and which we utilize in this article, include
  \texttt{metis}, \texttt{smt}, \texttt{simp}, \texttt{blast},
  \texttt{force}, and \texttt{auto}. In fact, in each case where those
  occur in the presented Isabelle/HOL formalizations, we have actually
  first used a generic hammer-tool, called sledgehammer
  \cite{sledgehammer}, which calls state-of-the-art ATPs to prove the
  statements in question fully automatically and without the need for
  specifying the particularly required premises; sledgehammer, in case
  of success, subsequently attempts to reconstruct the external proofs
  reported by the ATPs in Isabelle/HOL's trusted kernel by applying
  the mentioned automated reasoning tools.} In this part we conclude
from the premises and definitions (lines 5--8) that a Godlike being
possibly exists (theorem T3 in line 11):
$\lfloor \boldsymbol\Diamond \boldsymbol\exists^E \texttt{G}\rfloor$;
this follows from theorems T1 and T2 that are proved in lines 9 and
10. Note that, using binder notation,
$\lfloor \boldsymbol\Diamond \boldsymbol\exists^E \texttt{G}\rfloor$
can be more intuitively presented as
$\lfloor \boldsymbol\Diamond \boldsymbol\exists^E x.  \texttt{G}
x\rfloor$.
The most essential definition, the definition of property
$\texttt{G}$, which is of type $\gamma$ and which defines a Godlike
being $x_\texttt{e}$ to possess all (intensional!) positive properties
$\mathcal{P}$, is given in line 5. Premises that govern the notion of
(intensional) positive properties $\mathcal{P}$ are A1 (which is split
into A1a and A1b), A2 and A3; see lines 6--8. Scott \cite{ScottNotes}
actually avoids axiom A3 and instead directly postulates T2 (the sole
purpose of A3 is to support T2). Although we here explicitly include
the inference from A3 to T2, it could also be left out without any
implications for the rest of the proof.

Part II of the argument is presented in lines 12--20. In line~13 we
switch from base modal logic \textbf{K} to logic \textbf{KB} by postulating
symmetry of the accessibility relation \texttt{r}. Utilizing the same
tools as before, and by exploiting theorems T3, T4 and T5,
we finally prove, in line 20, the main theorem T6, which states that a
Godlike being necessarily exists:
$\lfloor \boldsymbol\Box \boldsymbol\exists^E \texttt{G}\rfloor$,
resp.~$\lfloor \boldsymbol\Box \boldsymbol\exists^E x.  \texttt{G}
x\rfloor$ using binder notation.

Consistency of the Isabelle/HOL theory \texttt{ScottVariant}, as
introduced up to here,
is confirmed by the model finder Nitpick \cite{Nitpick} in line 21 (which constructs a model with one
world and one Godlike entity).

In lines 23--29 modal collapse is proved. This is one of the rare cases in
our experiments where direct proof automation with Isabelle/HOL's integrated automated reasoning tools (incl. sledgehammer \cite{sledgehammer}) still fails. A
little interactive help is needed here to show that modal collapse
indeed follows from the premises in Scott's variant of G\"odel's
argument.

For more background information and details on the formalization  of Scott's argument,
and also on the arguments by Anderson and Fitting as presented in the following
sections, we refer to Fitting's book~\cite[\S 11]{fitting02:_types_tableaus_god} and our previous work \cite{C65}.

\subsection{Positive Properties and Ultrafilters: Scott}
Interesting  findings  regarding positive properties and ultrafilters
in Scott's variant are   revealed  in Fig.~\ref{fig:Scott2}. 

Theorem U1, which is proved in lines 32--37, states that the set of
positive properties $\mathcal{P}$
in Scott's variant  constitutes a $\gamma$-Ultrafilter. 

In line 38, a modified notion of positive properties $\mathcal{P'}$ is defined as the set of
properties $\varphi$ whose rigidly intensionalized extensions
${\boldsymbol{\downarrow}}\varphi$ are in
$\mathcal{P}$. It is then shown in theorem U2 (lines 39--44), that also
$\mathcal{P'}$ constitutes a $\gamma$-Ultrafilter. And theorem U3 in line 45
shows that these two sets, $\mathcal{P}$ and $\mathcal{P'}$, are
in fact equal.
 
In line 47 we switch from logic \textbf{KB} to logic \textbf{S5} by
postulating reflexivity and transitivity of the accessibility relation
\texttt{r} in addition to symmetry (line 13 in Fig.~\ref{fig:Scott1}); and we show consistency again (line 48). In the remaining lines 49--53 in Fig.~\ref{fig:Scott2} we show that the Barcan and the converse Barcan formulas are valid for types
$\texttt{e}$ and $\gamma$; we use for the former
type actualist quantifiers (as in the argument) and for the latter type possibilist quantifiers.

\section{Anderson's Variant of G\"odel's Argument} \label{Anderson}
\begin{figure}[tp] \centering \footnotesize
\fcolorbox{white}{gray!10}{
\begin{tabular}{ll} 
\multicolumn{2}{c}{\textbf{Anderson's Axioms and Definitions}} \\\hline \\[-.5em]
(df.$\texttt{G}^A$) & $\texttt{G}^A\, x \equiv \forall Y_\gamma. \mathcal{P}\, Y
  \leftrightarrow \Box (Y x)$ \\ 
(A1a) & $\forall X. \mathcal{P}({\rightharpoondown} X)  \rightarrow \neg (
  \mathcal{P}\, X)$     \qquad {where $\rightharpoondown$ is
    set/predicate negation} \\
(A2) & $\forall X Y. (\mathcal{P}\, X \wedge \Box  (\forall^E
  z. X z \rightarrow Y z)) 
  \rightarrow  \mathcal{P}\, Y$ \\
(T2) & $\mathcal{P}\, \texttt{G}^A$ \\
(A4) & $\forall X. \mathcal{P}\, X \rightarrow  \Box (\mathcal{P}\,
  X)$ \\
(df.$\mathcal{E}^A$) & $\mathcal{E}^A\, Y x \equiv \forall
  Z. \Box (Z x) \leftrightarrow \Box  (\forall^E
  z. Y z \rightarrow Z z)$ \\
(df.$\text{NE}^A$) & $\text{NE}^A\, x \equiv \forall Y.  \mathcal{E}^A\, Y
  x \rightarrow \Box \exists^E\, Y$ \\
(A5) &  $\mathcal{P}\, \text{NE}^A$ \\  
\end{tabular}
}
\vskip1em
\includegraphics[width=1\textwidth]{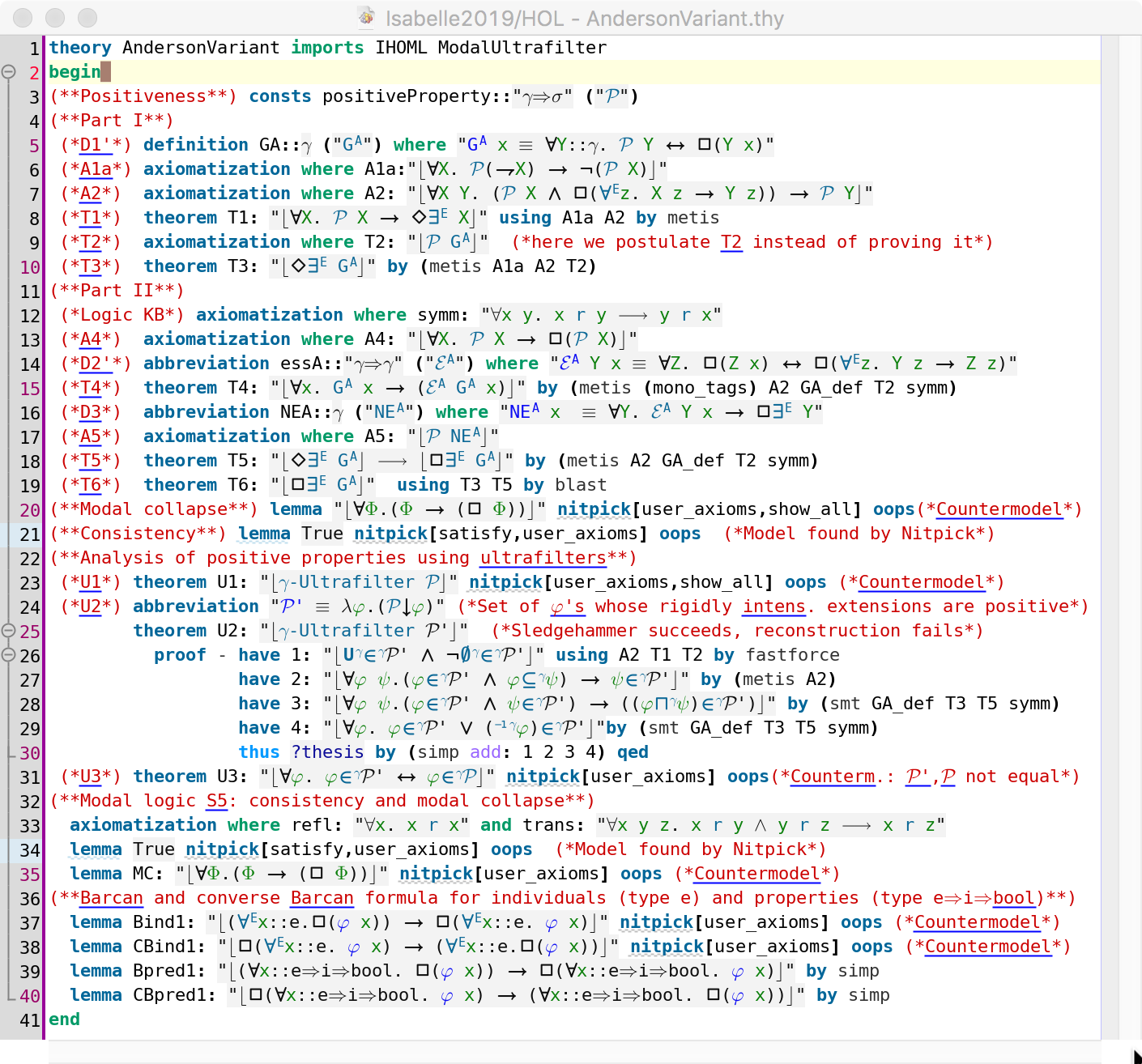}
\caption{Anderson's variant of G\"odel's argument, following Fitting \cite{fitting02:_types_tableaus_god}. \label{fig:Anderson}}
\end{figure}

Anderson's variant of G\"odel's argument is presented in
Fig.~\ref{fig:Anderson}. 

A central change in comparison to Scott's variant concerns 
Scott's premises A1a and A1b. Anderson drops A1b and only keeps A1a:
``If a property is positive, then its negation is not positive''.
This modification, however, has the effect that the
necessary existence of a Godlike being would no longer follow (and
the reasoning tools in Isabelle/HOL can confirm this; not shown
here). Anderson's variant
therefore introduces further emendations: it strengthens the notions of Godlikeness (in line 5) and
essence (in line 14). The emended notions, referred to by $\texttt{G}^A$ and
$\mathcal{E}^A$, are as follows:
\begin{description}
  \item[$\texttt{G}^A$] An individual $x$ is Godlike $\texttt{G}^A$ if and
    only if all and only the necessary/essential properties of $x$ are
    positive, i.e., $\texttt{G}^Ax \boldsymbol\equiv \boldsymbol\forall Y (\mathcal{P}\, Y  \boldsymbol\leftrightarrow \Box (Y x))$.
  \item[$\mathcal{E}^A$] A property $Y$ is an essence $\mathcal{E}^A$ of an individual
    $x$ if and only if all of $x$'s necessary/essential properties are
    entailed by $Y$ and (conversely) all properties entailed by $Y$
    are necessary/essential properties of $x$.
\end{description}
As is shown in lines 3--19, no further modifications are required
to ensure that the intended theorem T6, the necessary existence of a
$\texttt{G}^A$-like being, can (again) be proved.\footnote{In a very
  stringent interpretation this statement is not entirely true: Theorem
  T2 in Scott's argument, which was derived in Fig.~\ref{fig:Scott1}
  from axiom A3 and the definition of $\texttt{G}$, is now directly
  postulated here (for simplicity reasons) and axiom A3, which had no
  other purpose besides supporting T2, is
  dropped. This simplification, however, is obviously independent from the
  aspects as discussed.}

In line 20, the model finder Nitpick confirms that modal collapse is
indeed countersatisfiable in Anderson's variant of G\"odel's
argument. As expected, the reported countermodel consists of two
worlds and one entity.

Consistency of theory \texttt{AndersonVariant} is confirmed by Nitpick
in line 21, by finding a model with only one world and one entity (not shown).

\subsection{Positive Properties and Ultrafilters: Anderson}
Regarding positive properties and ultrafilters an interesting
difference to our prior observations for Scott's version is revealed
by the automated reasoning tools: the set of positive properties
$\mathcal{P}$ in Anderson's variant does \emph{not} constitute a
$\gamma$-Ultrafilter; Nitpick finds a countermodel to statement U1 in
line 23 that consists of two worlds and one entity.  However, the
modified notion $\mathcal{P}'$, i.e., the set of all properties
$\varphi$, whose rigidly intensionalized extensions are in
$\mathcal{P}$ (line 24), still is a $\gamma$-Ultrafilter; see theorem
U2, which is proved in lines 25--30.  Consequently, the sets
$\mathcal{P}$ and $\mathcal{P}'$ are not generally equal anymore and
Nitpick reports a countermodel for statement U3 in line 31.

In lines 32--40, we once again switch from logic \textbf{KB} to logic \textbf{S5}, we again show
consistency, and we again analyze the Barcan and the converse Barcan
formulas for types $\texttt{e}$ and $\gamma$.  In contrast to before,
the Barcan and converse Barcan formulas for type $\texttt{e}$, when
formulated with actualist quantifiers, are not valid anymore; Nitpick
presents countermodels with two worlds and two entities.

\section{Fitting's Variant of G\"odel's Argument} \label{Fitting}
\begin{figure}[tp] \centering \footnotesize
\fcolorbox{white}{gray!10}{
\begin{tabular}{ll} 
\multicolumn{2}{c}{\textbf{Fitting's Axioms and Definitions}} \\\hline \\[-.5em]
(df.$\texttt{G}$) & $\texttt{G}\, x \equiv \forall Y_\delta. \mathcal{P}\, Y
  \rightarrow \llparenthesis Y x\rrparenthesis$ \\ 
(A1a) & $\forall X. \mathcal{P}({\rightharpoondown} X)  \rightarrow \neg (
  \mathcal{P}\, X)$     \qquad {where $\rightharpoondown$ is
    set/predicate negation} \\
(A1b) & $\forall X. \neg ( \mathcal{P}\, X)\rightarrow
        \mathcal{P}({\rightharpoondown} X)$ \\
(A2) & $\forall X Y. (\mathcal{P}\, X \wedge \Box  (\forall^E
  z. \llparenthesis X z \rrparenthesis \rightarrow \llparenthesis Y z \rrparenthesis)) 
  \rightarrow  \mathcal{P}\, Y$ \\
(T2) & $\mathcal{P}{\downarrow}\texttt{G}$ \\
(df.$\mathcal{E}$) & $\mathcal{E}\, Y x \equiv \llparenthesis Y x \rrparenthesis \wedge (\forall
  Z. \llparenthesis Z x \rrparenthesis \rightarrow \Box  (\forall^E
  z. \llparenthesis Y z \rrparenthesis \rightarrow \llparenthesis Z z \rrparenthesis)$ \\
(df.$\text{NE}$) & $\text{NE } x \equiv \forall Y.  \mathcal{E}\, Y
  x \rightarrow \Box (\exists^E\, z. \llparenthesis Y z \rrparenthesis)$ \\
(A5) &  $\mathcal{P}{\downarrow}\text{NE}$ \\ 
\end{tabular}
}
\vskip1em
\includegraphics[width=1\textwidth]{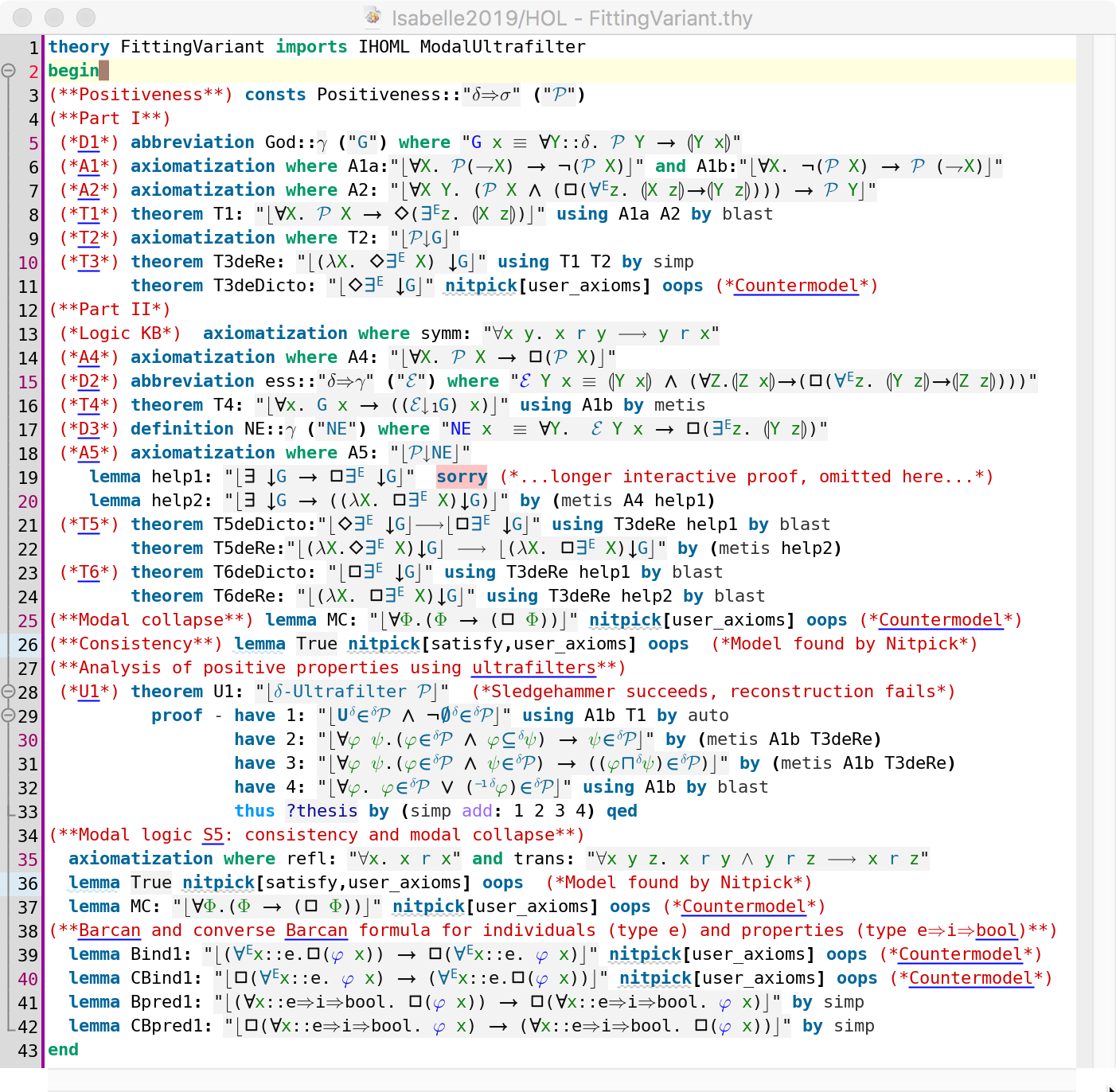}
\caption{Fitting's variant of G\"odel's argument. \label{fig:Fitting}}
\end{figure}
In Fitting's variant of G\"odel's Argument, see Fig.~\ref{fig:Fitting},
the notion of positive properties $\mathcal{P}$ in the definition of
Godlikeness $\texttt{G}$ ranges over extensions of properties, i.e.,
over terms of type $\delta$, and not over $\gamma$-type intensional
properties as in Scott's and Anderson's variants. In
Fitting's understanding, positive properties are thus fixed from world
to world, while they are world-dependent in Scott's and Anderson's. In
technical terms, Scott (resp.~G\"odel) defines $\texttt{G}\, x$ as 
$\boldsymbol\forall Y_\gamma. \mathcal{P} Y
\boldsymbol\rightarrow Y x$
(line 5 in Fig.~\ref{fig:Scott1}), whereas Fitting modifies this into
$\forall Y_\delta. \mathcal{P} Y
\boldsymbol\rightarrow \llparenthesis Y x \rrparenthesis$
(line 5 in Fig.~\ref{fig:Fitting}).  In an analogous way, the notion
of essence is emended by Fitting: in Scott's variant, see line
15 in Fig.~\ref{fig:Scott1}, $\mathcal{E}\, Y x$ is defined as
$Y x \wedge (\forall
  Z. Z x \rightarrow \Box  (\forall^E
  z. Y z \rightarrow Z z))$, while it becomes
$\llparenthesis Y x \rrparenthesis \wedge (\forall
  Z. \llparenthesis Z x \rrparenthesis \rightarrow \Box  (\forall^E
  z. \llparenthesis Y z \rrparenthesis \rightarrow \llparenthesis Z z \rrparenthesis)$ in Fitting's variant (see line 15 in Fig.~\ref{fig:Fitting}).  

The definition of necessary
existence \text{NE} in
line 17 is adapted accordingly, and in several other places of
Fitting's variant respective emendations are required to suitably
address his alternative interpretation of G\"odel's notion of positive
properties (see, e.g., theorem T2 in line 9 or axiom A5 in line
18). Fitting's expressive logical system (IHOML) also allows us to distinguish between \textit{de dicto} and \textit{de re} readings of
theorems T3, T5, and T6. Except for the \textit{de dicto} reading of T3, which has a countermodel
with two worlds and two entities, all of these statements are proved
automatically by the reasoning tools integrated with Isabelle/HOL.

As intended by Fitting, modal collapse is not provable anymore, which
can be seen in line 25, where Nitpick reports a countermodel with two
worlds and one entity. 

Consistency of the Isabelle/HOL theory
\texttt{FittingVariant}, as introduced up to here, is confirmed by Nitpick in line 26 (one world,
one entity). 

\subsection{Positive Properties and Ultrafilters: Fitting}
The type of $\mathcal{P}$ has
changed in Fitting's variant from the prior $\gamma{\Rightarrow}\sigma$ to
$\delta{\Rightarrow}\sigma$.  Hence, in our ultrafilter analysis, the
notion of a $\gamma$-Ultrafilter no longer applies and we must consult
the corresponding notion of a $\delta$-Ultrafilter. Theorem U1, which
is proved in lines 28--33 of Fig.~\ref{fig:Fitting}, confirms that Fitting's
emended notion of $\mathcal{P}$ indeed constitutes a
$\delta$-Ultrafilter.

In line 35 we again switch from modal logic \textbf{KB} to logic \textbf{S5}.
Consistency of the Isabelle/HOL theory \texttt{FittingVariant} in S5 is confirmed
in line 36, and countersatisfiability of modal collapse is reconfirmed
in line 37. 

Moreover, like for Anderson's variant before, we get a countermodel for 
the Barcan formula and the converse Barcan formula on type
$\texttt{e}$, when formulated with actualist quantifiers. The Barcan
formula and its converse are proved valid for type $\gamma$.

\section{Conclusion}
Anderson and Fitting both succeed in altering G\"odel's modal
ontological argument in such a way that the intended result, the
necessary existence of a Godlike being, is maintained while modal
collapse is avoided. And both solutions, from a cursory reading, are
quite different.

We conclude by rephrasing in more precise, technical terms what has
been mentioned at abstract level already in the mentioned related
article~\cite[\S 2.3]{J47}:

In order to compare the argument variants by Scott, Anderson, and
Fitting, two notions of ultrafilters were formalized in Isabelle/HOL:
A $\delta$-Ultrafilter, of type
$(\delta{\Rightarrow}\sigma){\Rightarrow}\sigma$, is defined on the
powerset of individuals, i.e., on the set of rigid properties, and a
$\gamma$-Ultrafilter, which is of type
$(\delta{\Rightarrow}\sigma){\Rightarrow}\sigma$, is defined on the
powerset of concepts, i.e., on the set of non-rigid, world-dependent
properties. In our formalizations of the variants, a careful
distinction was made between the original notion of a positive
property $\mathcal{P}$ that applies to (intensional) properties and a
restricted notion $\mathcal{P'}$ that applies to properties
whose rigidified extensions are $\mathcal{P}$-positive. Using these definitions
the following results were proved computationally:
\begin{itemize}
    \item In Scott's variant both $\mathcal{P}$ and $\mathcal{P'}$
      coincide, and both are $\gamma$-Ultrafilters.
    \item In Anderson's variant $\mathcal{P}$ and $\mathcal{P'}$ do
      not coincide, and only $\mathcal{P'}$, but not $\mathcal{P}$, is a
      $\gamma$-Ultrafilter.
    \item In Fitting's variant, the $\mathcal{P}$ in the sense of
      Scott and Anderson is not considered an appropriate
      notion. However, Fitting's emended notion of a positive property
      $\mathcal{P}$, which applies to extensions of properties,
      corresponds to our definition of $\mathcal{P'}$ in Scott's and
      Anderson's variants; and, as was to be expected, Fitting's emended notion
      of $\mathcal{P}$ constitutes a $\delta$-Ultrafilter.
\end{itemize}
The presented computational experiments thus reveal an intriguing correspondence
between the variants of the ontological argument by Anderson and
Fitting, which otherwise seem quite different. The variants of
Anderson and Fitting require that only the restricted notion of a
positive property is an ultrafilter.

The notion of positive properties in G\"odel's ontological argument is
thus aligned with the mathematical notion of a (principal) \textit{modal} ultrafilter
on intensional properties, and to avoid modal collapse it is
sufficient to restrict the modal ultrafilter-criterion to property
extensions. In a sense, the notion of Godlike being ``$\texttt{G} x$'' of
G\"odel is thus in close correspondence to the $x$-object in a principal
modal ultrafilter ``$F_x$'' of positive properties. This appears interesting
and relevant, since metaphysical existence of a Godlike being is now
linked to existence of an abstract object in a mathematical theory. 

Further research could look into a formal analysis of monotheism and
polytheism for the studied variants of G\"odel's ontological
argument. We conjecture that different notions of equality will
eventually support both views, and a respective formal exploration
study could take Kordula \'Swi\c etor\-zecka's related
work~\cite{Swietorzecka19} as a starting point.

\paragraph{Acknowledgements.} This work was supported by
VolkswagenStiftung under grant CRAP (Consistent Rational Argumentation
in Politics). As already mentioned, the technical results 
presented in this article have been summarized at abstract level in a joint article with
Daniel Kirchner and Ed Zalta.  We are also grateful to the anonymous reviewer.

\bibliographystyle{plain}

\end{document}